\documentclass[aps,prl,reprint,amsmath,amssymb,graphicx,longbibliography]{revtex4-2}

\usepackage{bm}

\usepackage[normalem]{ulem}
\usepackage[colorlinks=true,allcolors=blue,bookmarks=false,pdfusetitle]{hyperref}
\usepackage{url}
\usepackage{xcolor} 
\usepackage{graphicx}
\usepackage{breqn}
\usepackage{braket}

\usepackage{tikz}

\makeatletter
\let\cat@comma@active\@empty
\makeatother

\begin{document}

\title{Maxwell Relation between Entropy and Atom-Atom Pair Correlation}

\author{Raymon S. Watson}
\author{Caleb Coleman}
\author{Karen V. Kheruntsyan}
\affiliation{School of Mathematics and Physics, University of Queensland, Brisbane,  Queensland 4072, Australia}

\date{\today{}}

\begin{abstract}
\noindent 
For many-particle systems with short-range interactions the local (same point) particle-particle pair correlation function represents a thermodynamic quantity that can be calculated using the Hellmann-Feynman theorem. Here we exploit this property to derive a thermodynamic Maxwell relation between the local pair correlation and the entropy of an ultracold Bose gas in one dimension (1D). To demonstrate the utility of this Maxwell relation, we apply it to the computational formalism of the stochastic projected Gross-Pitaevskii equation (SPGPE) to determine the entropy of a finite-temperature 1D Bose gas from its atom-atom pair correlation function. Such a correlation function is easy to compute numerically within the SPGPE and other formalisms, which is unlike computing the entropy itself. Our calculations can be viewed as a numerical experiment that serves as a proof-of-principle demonstration of an experimental method to deduce the entropy of a quantum gas from the measured atom-atom correlations. 
\end{abstract}

\maketitle

\emph{Introduction.}---Entropy plays a fundamental role in thermodynamics, statistical mechanics, and quantum information theory. However, measuring it directly or calculating it from the defining multiplicity function or the density matrix of an interacting many-body system often represents a formidable challenge. Instead, the entropy is often deduced from other thermodynamic quantities (such as the heat capacity or the free energy) using the relevant thermodynamic relations \cite{Huang_book,Callen_book}. Here, we derive and discuss a thermodynamic Maxwell relation by which the entropy of a quantum many-body system with short-range interactions can instead be related to, and hence deduced from, the local particle-particle correlation function. Such a pair correlation function characterizes the probability of two particles to be found in the same position compared to uncorrelated particles and can often be computed 
using methods of many-body and quantum field theory either analytically or numerically \cite{Gangardt2003,Kheruntsyan2003}.
It can also be measured experimentally in, e.g., ultracold quantum gas experiments using photoassociation \cite{Weiss_g2_2005,Hulet_g2_2005}.

The surprising aspect of the Maxwell relation between the pair correlation 
and the entropy that we discuss here is that the pair correlation function is usually viewed and treated as a typical two-body observable, whereas the entropy is a macroscopic thermodynamic quantity. However, what promotes the pair correlation into a thermodynamic quantity as well is the fact that we are only considering many-body systems with short-range interactions that can be characterized by the $s$-wave scattering length \cite{Pethick_Smith_book,Pitaevskii_Stringari_book}. In this case, the interparticle interactions can be approximated by a simple contact interaction, meaning that the two-body correlation function at zero interparticle separation indeed becomes a thermodynamic quantity. This was first demonstrated by Lieb and Liniger in their seminal work on the exact Bethe ansatz treatment of a uniform one-dimensional (1D) Bose gas with repulsive contact ($\delta$-function) interactions \cite{Lieb-Liniger-I,*Lieb-Liniger-II}. By using the Hellmann-Feynman theorem and differentiating the total ground state (zero-temperature, $T\!=\!0$) energy of the gas with respect to the interaction strength, Lieb and Liniger were able to calculate the mean interaction energy component, which itself is proportional to the unnormalized pair correlation function (see below).

The extension of the Hellmann-Feynman theorem to finite-temperature systems \cite{Kheruntsyan2003,Kheruntsyan2005,PhysRevB.108.L140403}, together with the exact Yang-Yang thermodynamic Bethe ansatz (TBA) solution for the 1D Bose at finite temperature \cite{Yang-Yang}, was later utilized to calculate the local pair correlation at any temperature and interaction strength. In this case, the pair correlation function is related to the partial derivative of the Helmholtz free energy with respect to the interaction strength \cite{Kheruntsyan2003,Kheruntsyan2005,Kerr2024F}. In this Letter, we take this relationship a step further by combining it with the fact that the partial derivative of the same Helmholtz free energy with respect to the temperature, on the other hand, gives the entropy of the system according to the canonical ensemble formalism of statistical mechanics. Therefore, by using the commutative property of mixed second derivatives of the Helmholtz free energy (with respect to the interaction strength and temperature) one obtains the Maxwell relation between the pair correlation and the entropy that we discuss here. 

As a practical application of this Maxwell relation, we utilize it  for computing the entropy of a weakly interacting 1D Bose gas in the quasicondensate regime in the context of the classical $c$-field approach of the stochastic projected Gross-Pitaevskii equation (SPGPE) \cite{castin2000,gardiner2002stochastic,gardiner2003stochastic,blakie2008dynamics,spgpe_Rooney_2012,Bradley_2005}. The SPGPE is a well established and widely used numerical approach for computing thermal equilibrium and dynamical properties of finite-temperature Bose gases, such as partially condensed Bose-Einstein condensates in 2D and 3D (see, e.g., \cite{blakie2008dynamics,spgpe_Rooney_2012,Bradley2015} and references therein), or phase-fluctuating quasicondensates in 1D 
\cite{castin2000,Sinatra_PRL_2001,Bradley_2005,blakie2008dynamics,stimming2010,Grisins2011,stimming2011,Bradley2015,Bouchoule2016,Deuar2018,Thomas2021,Bayocboc2022,Bayocboc2023}. Despite its wide applicability to ultracold quantum gas systems, computing the entropy of such systems within the SPGPE has not been accomplished prior to this work. Here, we compute the entropy of a 1D quasicondensate within the SPGPE approach; we restrict ourselves to the 1D Bose gas because of the availability of the exact TBA solution for both the entropy and the pair correlation function, to which we compare, and hence validate, our numerical SPGPE results. However, we point out that the Maxwell relation derived and discussed here is equally applicable to 2D and 3D systems, as well as to Fermi gas systems with similar contact interactions, such as the Yang-Gaudin model in 1D \cite{Zwerger_Tan_2011} or a two-component 3D Fermi gas near the BCS-BEC crossover \cite{Tan_contact_2008,Werner2009,Pitaevskii_Stringari_book,Cherny2021}.

\emph{Lieb-Liniger model and two-particle correlation.}---We start by considering the Lieb-Liniger model describing a uniform 1D gas of $N$ bosons of mass $m$ interacting via a pairwise $\delta$-function potential on a line of length $L$ with periodic boundary conditions and of linear density of $n=N/L$. In second-quantized form, the Hamiltonian of such a system is given by
\begin{align}
	\hat{H}	= & -\frac{\hbar^{2}}{2m}  \int dx\,  \hat{\Psi}^{\dagger} \frac{\partial^2 \hat{\Psi}}{\partial x^2} + \frac{\chi}{2} \int dx\, \hat{\Psi}^{\dagger} \hat{\Psi}^{\dagger} \hat{\Psi} \hat{\Psi}.\label{eq:H}
\end{align}
Here, $\hat{\Psi}^{\dagger}(x)$ and $\hat{\Psi}(x)$ are the bosonic field creation and annihilation operators, whereas $\chi$ quantifies the strength of boson-boson interactions, assumed to be repulsive ($\chi>0$). This interaction strength can be expressed in terms of the 3D $s$-wave scattering length $a$ via $\chi\approx2\hbar\omega_{\perp}a$ \cite{Olshanii1998}, away from a confinement induced resonance, where $\omega_{\perp}$ is the frequency of the harmonic potential in the transverse (tightly confined) dimension.

The normalized two-point particle-particle correlation  is  defined in terms of the field operators as the expectation value of a normally ordered product of two density operators, $\hat{n}(x)=\hat{\Psi}^\dagger(x)\hat{\Psi}(x)$ and $\hat{n}(x')=\hat{\Psi}^\dagger(x')\hat{\Psi}(x')$: 
\begin{equation}
    g^{(2)}(x,x') = \frac{\langle \hat{\Psi}^\dagger(x)\hat{\Psi}^\dagger(x')\hat{\Psi}(x')\hat{\Psi}(x) \rangle}{n(x)n(x')}.
\end{equation}
In other words, the pair correlation $g^{(2)}(x,x')$ is a normalized and normally ordered density-density correlation function. It is normalized to the product of mean densities $n(x)=\langle \hat{n}(x)\rangle$ and $n(x')=\langle \hat{n}(x')\rangle$ at points $x$ and $x'$ so that for uncorrelated particles (with $\langle \hat{\Psi}^\dagger(x)\hat{\Psi}^\dagger(x')\hat{\Psi}(x')\hat{\Psi}(x) \rangle=\langle \hat{\Psi}^\dagger(x)\hat{\Psi}(x) \rangle \langle \hat{\Psi}^\dagger(x')\hat{\Psi}(x')\rangle$), one has $g^{(2)}(x,x')=1$. For values of $g^{(2)}(x,x')\neq 1$, the pair correlation characterizes an enhanced [$g^{(2)}(x,x')> 1$] or suppressed [$g^{(2)}(x,x')< 1$] probability of finding two particles at positions $x$ and $x'$, respectively, compared to uncorrelated particles.

Because of the translational invariance of the uniform system that we are considering, where $n(x')\!=\!n(x)\!=\!n$, the above pair correlation $g^{(2)}(x,x')$ can only depend on the relative distance $|x-x'|$ between the two particles, i.e., $g^{(2)}(x,x')\!=\!g^{(2)}(|x-x'|)$.  The local or the same-point ($x\!=\!x'$) correlation then corresponds to
\begin{equation}\label{eq:g2-definition}
    g^{(2)} \equiv g^{(2)}(0)= \frac{\langle \hat{\Psi}^{\dagger}(x)\hat{\Psi}^{\dagger}(x) \hat{\Psi}(x)\hat{\Psi}(x)\rangle }{n^2}.
\end{equation}

In the canonical formalism, the partition function $Z(T,N,L,\chi)$ can be written in terms of either the Helmholtz free energy $F$ or the Hamiltonian $\hat{H}$ via $Z\!=\! \exp(-F/k_BT) = \text{Tr}\exp(-\hat{H}/k_BT)$. By differentiating the Helmholtz free energy $F(T,L,N,\chi)\!=\!-k_BT\ln Z$ with respect to the interaction strength $\chi$, at constant $N$, $L$, and $T$, one finds that \cite{Kheruntsyan2003}
\begin{equation}
 \left(\frac{\partial F}{\partial \chi}\right)_{T,L,N} \!\!= \frac{1}{Z} \text{Tr}\left(e^{-\hat{H}/k_BT} \frac{\partial \hat{H}}{\partial \chi}\right) = 
 \frac{1}{2}\overline{G^{(2)}},
 \label{eq:G2}
 \end{equation}
where we have introduced an integrated unnormalized correlation function $\overline{G^{(2)}}\equiv\int dx \langle \hat{\Psi}^{\dagger}\hat{\Psi}^{\dagger}\hat{\Psi}\hat{\Psi}\rangle$. Since $\overline{G^{(2)}}=L\langle \hat{\Psi}^{\dagger}\hat{\Psi}^{\dagger}\hat{\Psi}\hat{\Psi}\rangle=Ln^2g^{(2)}$ for a uniform system, Eq.~\eqref{eq:G2} can be rewritten as
\begin{equation}\label{eq:g2-canonical}
    g^{(2)} = \frac{2}{Ln^2} \left(\frac{\partial F}{\partial \chi}\right)_{T,L,N}. 
\end{equation}
This relationship between the local pair correlation and the Helmholtz free energy is what was used in Ref. \cite{Kheruntsyan2003} to calculate the $g^{(2)}$ function using the exact Yang-Yang TBA \cite{Yang-Yang} solution for $F$, as a function of the dimensionless interaction strength $\gamma$ and the dimensionless temperature $\tau$, defined, respectively, via:
\begin{equation}
    \gamma = \frac{m\chi}{\hbar^2n},\qquad \tau = \frac{2mk_BT}{\hbar^2 n^2}.
\end{equation}
We note here that these two dimensionless parameters completely characterize the thermodynamic properties of a uniform 1D Bose gas \cite{Yang-Yang, Kheruntsyan2003}.

\emph{Maxwell relation.}---We now recall that the partial derivative of the same Helmholtz free energy with respect to temperature $T$ in the canonical formalism gives the entropy $S=S(T,L,N,\chi)$ of the system:
\begin{equation}
S=-\left(\frac{\partial F}{\partial T}\right)_{L,N,\chi}.
\label{eq:entropy}
\end{equation}

Combining  Eqs.~\eqref{eq:G2} and \eqref{eq:entropy} with the commutative property of mixed second derivatives of $F$, i.e. $\frac{\partial}{\partial \chi}\left( \frac{\partial F}{\partial T}\right)_{L,N} = \frac{\partial}{\partial T}\left( \frac{\partial F}{\partial \chi}\right)_{L,N} $, leads to the following Maxwell relation: 
\begin{equation}
\left(\frac{\partial S}{\partial \chi} \right)_{T,L,N}\!=-\frac{1}{2}\left(\frac{\partial \overline{G^{(2)}}}{\partial T}\right)_{L,N,\chi},
\label{eq:Maxwell}
\end{equation}
which for a uniform system can be rewritten as $\left(\frac{\partial S}{\partial \chi} \right)_{T,L,N}\!=- \frac{Ln^2}{2}\left( \frac{\partial g^{(2)}}{\partial T}\right)_{L,N,\chi}$.

Equation \eqref{eq:Maxwell} is one of this Letter's key results (see also Appendix A) and implies that the entropy of the gas at a specific value of $\chi$ (and some fixed values of $T$, $L$, and $N$) can be calculated by integrating the partial derivative of $\overline{G^{(2)}}$ with respect to $T$ over the interaction strength:
\begin{align}
S(T&,L,N,\chi)=S(T,L,N,\chi_0)\nonumber \\
&-\frac{1}{2}\int_{\chi_0}^{\chi}\left( \frac{\partial \overline{G^{(2)}}(T,L,N,\chi')}{\partial T}\right)_{L,N,\chi'}d\chi'.
\label{eq:entropy_from_integral}
\end{align}
Here, $S(T,L,N,\chi_0)$ serves the role of the integration constant and is assumed to be known for the method to work; in practice, it can be chosen to correspond to the entropy of an ideal ($\chi_0\!=\!0$) Bose gas (IBG), $S_{\mathrm{IBG}}\!=\!S(T,L,N,0)$, which can indeed be calculated for any $T$ using standard methods of statistical mechanics  \cite{Huang_book,kerr2024analytic}.

As a simple analytic illustration of the utility of Eq.~\eqref{eq:entropy_from_integral}, we calculate the entropy of a 1D Bose gas in highly degenerate, nearly ideal Bose gas regime that can be treated using perturbation theory with respect to $\gamma$ (see the results for the so-called decoherent quantum regime in Refs. \cite{Kheruntsyan2003,kerr2024analytic}, valid in the region $2\sqrt{\gamma}\ll \tau \ll 1$). In this regime, the normalized local pair correlation function $g^{(2)}$, which was calculated in Ref. \cite{Kheruntsyan2003} without resorting to the Helmholtz free energy, is given by
\begin{equation}
g^{(2)}=2-4\gamma/\tau^2.
\label{g2_III}
\end{equation}
Therefore, Eq.~\eqref{eq:entropy_from_integral} yields the following result for the corresponding entropy:
\begin{equation}
S=S_{\text{IBG}}-4k_BN\gamma^2/\tau^3.
\label{S_III}
\end{equation}


\emph{Application to the SPGPE approach.}---We now illustrate the utility of Eq.~\eqref{eq:entropy_from_integral} using a numerically computed pair correlation function within the SPGPE approach. This itself can be viewed as a numerical experiment demonstrating how one can deduce the entropy of a quantum gas from the measured atom-atom correlations. 

The SPGPE approach (see Appendix B) is a classical field or $c$-field method for computing thermal equilibrium and dynamical properties of degenerate Bose gases 
\cite{Bouchoule2016}
at finite temperatures \cite{castin2000,Sinatra_PRL_2001,Bradley_2005,blakie2008dynamics,spgpe_Rooney_2012,stimming2010,Grisins2011,stimming2011,Bradley2015,Bouchoule2016,Deuar2018,Thomas2021,Bayocboc2022,Bayocboc2023}. Evolving the SPGPE from an arbitrary initial state, for a sufficiently long tim (such that the memory of the initial state is lost), samples thermal equilibrium configurations  of the system from the grand-canonical ensemble. These configurations are represented by stochastic realizations of the complex $c$-fields $\Psi_{\mathbf{C}}(x,t)$, which we will denote as $\Psi_{\mathbf{C}}(x)$ for thermal equilibrium states. In the SPGPE approach, the pair correlation function $g^{(2)}$ is computed according to:
\begin{equation}
g^{(2)}=\frac{\langle \Psi_{\mathbf{C}}^{*}(x)\Psi_{\mathbf{C}}^{*}(x) \Psi_{\mathbf{C}}(x) \Psi_{\mathbf{C}}(x)\rangle}{\langle \Psi_{\mathbf{C}}^{*}(x) \Psi_{\mathbf{C}}(x) \rangle^2},
\label{g2-SPGPE}
\end{equation}
where the expectation values are over a large number of stochastic realizations.

\begin{figure}[t] 
    \includegraphics[width=8.4cm]{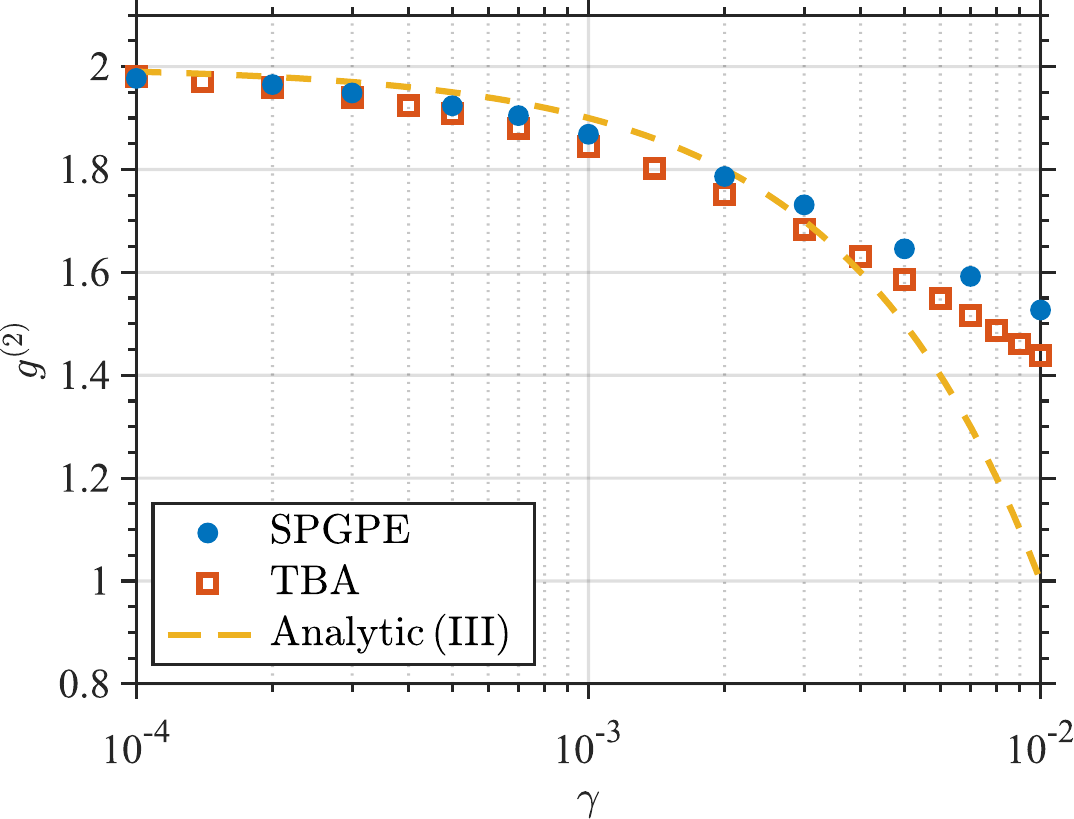}
    \caption{Normalized pair correlation $g^{(2)}$ for a 1D quasicondensate as a function of the dimensionless interaction strength $\gamma$, for a fixed dimensionless temperature $\tau=0.2$. The numerically computed data from the SPGPE simulations are shown as circles and are compared to the exact TBA result (squares), and the analytic approximation of Eq.~\eqref{g2_III} (dashes).}
    \label{fig:g2}
\end{figure}

We illustrate the utility of Eq. \eqref{eq:entropy_from_integral} for a 1D quasicondensate using a $c$-field approach within the regime of its applicability, which is restricted to the parameter range $\sqrt{\gamma}\!\ll\! \tau\! \ll\! 1$ \cite{castin2000,Bayocboc2023}. In Fig.~\ref{fig:g2} we first show the dependence of the normalized pair correlation function $g^{(2)}$ over a range of the dimensionless interaction strength $\gamma\in [10^{-4}, 10^{-2}]$, for a fixed value of the dimensionless temperature $\tau=0.2$, obtained from the SPGPE approach. For comparison, we also show the exact TBA result (squares) \cite{Kheruntsyan2003} and the approximate analytic result of Eq.~\eqref{g2_III} (dashed line).

As we see, in the limit of an ideal Bose gas ($\gamma \to 0$) at finite temperature, the pair correlation approaches the value of $g^{(2)}=2$, which is the Hanbury Brown--Twiss effect of bosonic bunching ($g^{(2)}>1$) first observed for photons from a chaotic (thermal) light source \cite{HBT_1956_Stellar,*HBT_1956_two_coherent,*HBT_1956_question} and more recently for an ultracold atomic gas above the transition to a Bose-Einstein condensate \cite{HBT_Esslinger_2005,HBT_BEC_Aspect_2005,Jeltes2007}. It corresponds to large density fluctuations and an enhanced probability of detecting two indistinguishable bosons in the same position due to the constructive interference of the respective probability amplitudes. As the strength of the repulsive interaction increases, the said probability decreases and manifests itself in the reduction of the value of $g^{(2)}$ below $2$ \cite{Kheruntsyan2003,Weiss_g2_2005}. At some finite, but still weak ($\gamma\! \ll\! 1$) interaction strength, the pair correlation crosses the coherent level of $g^{(2)}\!=\!1$ characteristic of a phase-fluctuating quasicondensate with suppressed density fluctuations, which itself shares the properties of a weakly interacting Bose-Einstein condensate in the mean-field description \cite{HBT_Esslinger_2005,HBT_BEC_Aspect_2005}. 

As the interaction strength increases further and approaches the regime of very strong or hard-core repulsion ($\gamma \to \infty$), also known as the Tonks-Girardeau limit of fermionization, the pair correlation reduces further down to $g^{(2)}\to 0$ (see Refs. \cite{Kheruntsyan2003,Weiss_g2_2005}). This reduction reflects the fact that the bosons are now strongly (anti)correlated and behave effectively as fermions, wherein the bosonic hard-core repulsion mimics the fermionic Pauli blocking. In the pair correlation function, such repulsion manifests itself as antibunching ($g^{(2)}<1$), which itself is due to the destructive interference of probability amplitudes for detecting two indistinguishable fermions in the same position \cite{Jeltes2007}. This regime, however, is beyond the applicability of the SPGPE approximation ($\gamma\ll \!1$, $2\gamma\!\ll\! \tau\!\ll \!1$; see, e.g., Refs.~\cite{Bayocboc2023,kerr2024analytic}, and references therein), and this is why in Fig.~\ref{fig:g2} we do not show the behavior of the $g^{(2)}$ beyond the weakly interacting regime of $\gamma \!\ll \!1$. Because of the same approximate nature of the SPGPE approach, we see that the SPGPE data for $g^{(2)}$, while agreeing well with the exact TBA results at small $\gamma$, start to deviate from the TBA results as $\gamma$ increases and approaches its upper bound of $\gamma=0.01$, where the condition $2\gamma\ll \tau$ is not well satisfied. Similarly, the analytic result of Eq.~\eqref{g2_III} deviates from TBA to a larger extent as $\gamma$ is increased, because it is applicable in an even more restricted region of $2\sqrt{\gamma}\!\ll\! \tau\ll \!1$ \cite{Kheruntsyan2003,kerr2024analytic}.

\begin{figure}[t] 
    \includegraphics[width=8.4cm]{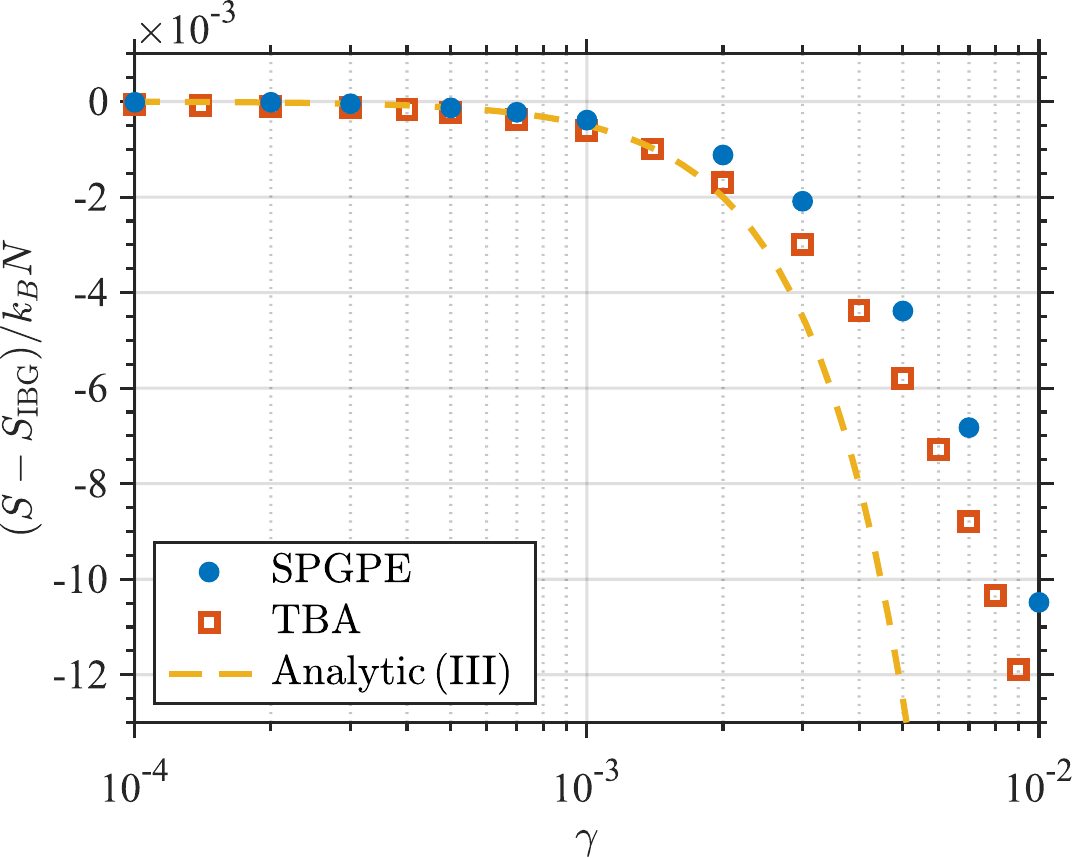}
    \caption{Entropy of a 1D quasicondensate as a function of the dimensionless interaction strength $\gamma$, relative to that of an ideal Bose gas at the same temperature, $S(\gamma,\tau)-S_{\text{IBG}}(\tau)$, for $\tau=0.2$. The SPGPE data computed using Eq.~\eqref{eq:entropy_from_integral} are shown as circles and are compared to the exact TBA data (squares), and the analytic approximation of Eq.~\eqref{S_III} (dashes).}
    \label{fig:S}
\end{figure}

We next deduce the entropy of the 1D quasicondensate using Eq.~\eqref{eq:entropy_from_integral} and the SPGPE data for $g^{(2)}$, except that now the integration in Eq.~\eqref{eq:entropy_from_integral}  is done numerically. To obtain the dependence of $S(\chi,T,L,N)$ on $\chi$ at a fixed $T$, or rather on the dimensionless $\gamma$ at a fixed $\tau$, we convert Eq.~\eqref{eq:entropy_from_integral} to the dimensionless units and first evaluate the derivative $\left(\partial g^{(2)}(\gamma',\tau)/\partial \tau\right)_{\gamma'}$ using the central difference scheme, for a range of values of $\gamma'$. We next evaluate the integral over $\gamma'$ numerically, as a function of the upper bound. The upper bound is scanned within $\gamma\in [10^{-4}, 10^{-2}]$, while fixing the lower bound at $\gamma_0=10^{-6}$, which is sufficiently low for the SPGPE results to be nearly identical to the IBG results at finite $T$, for which $g^{(2)}=2$ and $S(\gamma_0,T,L,N)\simeq S_{\text{IBG}}(T,L,N)$.
In Fig.~\ref{fig:S} we show the SPGPE result for the entropy difference per particle, $(S-S_{\text{IBG}})/k_BN$ obtained from the SPGPE approach as a function of $\gamma$, for a fixed value of the dimensionless temperature $\tau$. We again compare these data with the exact TBA result (squares) and the analytic result of Eq.~\eqref{S_III} (dashed line). As we see, the entropy is maximal in the ideal Bose gas limit ($\gamma \to 0$), where $g^{(2)}=2$ is also maximal, reflecting the large density fluctuations and excess randomness (bunching) in the probability of finding two indistinguishable bosons in the same position. As the strength of repulsive interactions increases, the random density fluctuations become more and more suppressed, which is also evident in the decrease of the entropy of the gas, as expected. Overall, we see a good agreement between the SPGPE and TBA results, particularly at small values of $\gamma$, where the condition of validity of the SPGPE approximation is better satisfied; the agreement becomes worse as $\gamma$ is increased, for the same reason as the discrepancy in the behavior of $g^{(2)}$ discussed earlier.

\emph{Summary and outlook.}---We have derived and discussed a new Maxwell relation by which the entropy of a quantum many-body system with contact two-body interactions can be related to, and deduced from, the local two-particle correlation function. We have validated this method though a numerical experiment based on the $c$-field SPGPE simulations and computed---for the first time (to the best of our knowledge) within the SPGPE formalism---the thermodynamic entropy of a weakly interacting 1D Bose gas in the quasicondensate regime.

The Maxwell relation derived here may find immediate applications, such as measuring the entropy and deducing the thermodynamic equation of state, in quantum gas experiments that take advantage of tunable interparticle interactions and measurements of atom-atom correlations using photoassociation or \emph{in situ} imaging techniques in quantum gas microscope setups. It can also be applied to other computational approaches, such as density matrix renormalization group and phase space stochastic gauge methods \cite{Drummond2004,Deuar2009}, that are capable of computing particle-particle correlations from the many-body wave function or density matrix formalism, but struggle to compute the entropy from, e.g., the multiplicity or the free energy. 

Apart from the 1D Bose gas model, the main results presented in this Letter, Eqs.~\eqref{eq:G2}, \eqref{eq:Maxwell} and \eqref{eq:entropy_from_integral}, can be easily extended to 2D and 3D Bose gas systems, Fermi gases, and Fermi and Bose gas mixtures. For example, for the Yang-Gaudin model of an interacting two-component Fermi gas in 1D \cite{Yang_Gaudin_1967,gaudin1967systeme,Zwerger_Tan_2011}, with the interaction Hamiltonian $\hat{H}_{\text{int}}\!=\!\chi \int \!dx  \hat{\Psi}_{\uparrow}^{\dagger} \hat{\Psi}_{\downarrow}^{\dagger} \hat{\Psi}_{\downarrow}\hat{\Psi}_{\uparrow}$, the results of Eqs.~\eqref{eq:G2}, \eqref{eq:Maxwell} and \eqref{eq:entropy_from_integral} continue to hold with the replacement $\overline{G^{(2)}}/2\!\to\! \overline{G_{\uparrow,\downarrow}^{(2)}}\!\equiv \!\int \!dx \langle \hat{\Psi}_{\uparrow}^{\dagger} \hat{\Psi}_{\downarrow}^{\dagger} \hat{\Psi}_{\downarrow}\hat{\Psi}_{\uparrow}\rangle$, where $\hat{\Psi}_{\uparrow}(x)$ and $\hat{\Psi}_{\downarrow}(x)$ are the fermionic field operators for the spin-up and spin-down components, respectively. Beyond quantum gas systems, these results can also aid the study of condensed matter systems that are characterized through the static structure factor which is measured in scattering experiments \cite{Vale_Bragg_2008,Hulet_2022}. Indeed, the static structure factor $S(\mathbf{k})$ is related to the nonlocal pair correlation $g^{(2)}(\mathbf{r})$ via a Fourier transform, $S(\mathbf{k})\!=\!1\!+\!n\int\! d\mathbf{r} g^{(2)}(\mathbf{r}) e^{-i\mathbf{k}\cdot \mathbf{r}}$. Therefore, a measurement or theoretical knowledge of $S(\mathbf{k})$ can be used to deduce the local pair correlation $g^{(2)}(0)\equiv g^{(2)}$ by an inverse Fourier transform, which can then be used for determining the thermodynamic entropy from Eq.~\eqref{eq:entropy_from_integral}. Finally, one can envisage derivation of related Maxwell relations for a large class of spin Hamiltonians \cite{Rist} in which: (a) the spin-spin interaction term can be similarly calculated using the Hellmann-Feynman theorem; and (b) the relevant spin-spin correlation function can be experimentally measured.

\begin{acknowledgments}
 \emph{Acknowledgement}---This work was supported through Australian Research Council Discovery Project Grant No. DP190101515.
 \end{acknowledgments}

~


\emph{\textbf{Appendix A: Thermodynamic potential behind the Maxwell equation for the atom-atom correlation function}}---In this appendix, we outline 
the fundamental thermodynamic identities behind the Maxwell relation \eqref{eq:Maxwell}. First, we note that in Eq.~\eqref{eq:G2}, which follows from the Hellmann-Feynman theorem, 
the integrated correlation function $\overline{G^{(2)}}=\int dx \langle \hat{\Psi}^{\dagger}(x)\hat{\Psi}^{\dagger}(x)\hat{\Psi}(x)\hat{\Psi}(x)\rangle$ (times a factor of $1/2$) can be viewed as an \emph{extensive} thermodynamic parameter \cite{extensive} that characterizes the variation of the system's internal energy $U=\langle \hat{H} \rangle$ with the interaction strength $\chi$ which itself is an \emph{intensive} parameter conjugate to $\overline{G^{(2)}}/2$.  (The factor of $1/2$ is an artifact of the conventional definition of the interaction part of the Hamiltonian \eqref{eq:H}, with $\langle \hat{H}_{\mathrm{int}}\rangle =\frac{\chi}{2}\overline{G^{(2)}}$, where $1/2$ can be either absorbed into the redefinition of the coupling constant, $\chi/2\to \chi$, or  kept as a multiplier in front of $\overline{G^{(2)}}$ whenever we talk about the integrated pair correlation as an extensive parameter.) The variation of the generalized Helmholtz free energy in the canonical formalism, $F=F(T,L,N,\chi)$, where for the 1D system the role of the volume $V$ is played by the length $L$, can therefore be written as  
\begin{equation}
dF=-SdT-PdL+\mu dN+(\overline{G^{(2)}}\!/2)d\chi,
\label{eq:dF}
\end{equation}
where $S=-(\partial F/\partial T)_{L,N,\chi}$ is the entropy, $P=-(\partial F/\partial L)_{T,N,\chi}$ is the pressure, $\mu =(\partial F/\partial N)_{T,L,\chi}$ is the chemical potential, and $\overline{G^{(2)}}/2=(\partial F/\partial \chi)_{T,L,N}$. From this, one can derive a set of Maxwell relations as usual, including the one between $S$ and $\overline{G^{(2)}}$, i.e., Eq.~\eqref{eq:Maxwell}. 

According to the standard formalism of thermodynamics (see, e.g., \cite{Callen_book,Alberty2002}), the fundamental equation \eqref{eq:dF} can be obtained via a Legendre transform,
\begin{equation}
F=U-TS+\chi (\overline{G^{(2)}}/2),
\end{equation}
from the Euler equation for the generalized internal energy of the system,
\begin{equation}
U=TS-PL+\mu N-\chi (\overline{G^{(2)}}\!/2),
\end{equation}
which is a function of only extensive parameters, $U=U(S,L,N,\overline{G^{(2)}}\!/2)$. The differential  of $U$ is given by  
\begin{equation}
dU=TdS-PdL+\mu dN-\chi d(\overline{G^{(2)}}\!/2),
\label{eq:dU}
\end{equation}
where $T\!=\!(\partial U/\partial S)_{L,N,\overline{G^{(2)}}}$, $P\!=\!-(\partial U /\partial L)_{S,N,\overline{G^{(2)}}}$, $\mu\!=\!(\partial U/\partial N)_{S,L,\overline{G^{(2)}}}$, and $\chi \!=\! -\left(\partial U/\partial(\overline{G^{(2)}}\!/2)\right)_{S,L,N}$ \cite{alt_chi}. We note that the negative sign in $\chi \!=\! -\left(\partial U/\partial(\overline{G^{(2)}}\!/2)\right)_{S,L,N}$ makes physical sense for positive $\chi$ (repulsive interactions) as the internal energy of the system increases when the pair correlation is decreased when approaching the `fermionized' regime of particle-particle antibunching where $g^{(2)}(0)\! \to\! 0$, as opposed to the weakly interacting regime where the gas displays bosonic bunching  $g^{(2)}(0) \!\to\! 2$ \cite{Kheruntsyan2003}. 

Furthermore, an equation similar to Eq.~\eqref{eq:dF} can be written down for the grand-canonical thermodynamic potential $\Omega=F-\mu N=U-TS+\chi (\overline{G^{(2)}}/2)-\mu N$, 
\begin{equation}
d\Omega=-SdT-PdL-Nd\mu+(\overline{G^{(2)}}\!/2)d\chi,
\label{eq:dO}
\end{equation}
with $\Omega=\Omega(T,L,\mu,\chi)$. Using additionally $\Omega=-PL$ for homogeneous systems, Eq.~\eqref{eq:dO} can be further rewritten as
\begin{equation}
LdP-SdT-Nd\mu+(\overline{G^{(2)}}\!/2)d\chi=0,
\label{eq:Gibb-Duhem}
\end{equation}
which takes the role of the generalized Gibbs-Duhem relation and implies, in particular, that among the four intensive parameters $\{P,T,\mu,\chi\}$ only three are independent; this, in turn, implies that the functional dependence of the fourth parameter on the other three takes the role of the thermodynamic equation of state, such as $P=P(T,\mu,\chi)$. For explicit examples of such equations of state for the uniform 1D Bose gas, see a recent review in Ref.~\cite{kerr2024analytic}.

We emphasize that all these generalizations of the thermodynamic relations, accounting for the changes of the interaction strength $\chi$, are applicable only to ultracold atomic gases in which the interactions are short ranged and can be accounted for via a single parameter ($\chi$), which itself can be varied via the $s$-wave scattering length $a$. These generalized thermodynamic relations can be adopted to describe short-range interacting Fermi gases \cite{Tan_contact_2008,Zwerger_Tan_2011,Braaten_Tan_review,Pitaevskii_Stringari_book,Cherny2021}, Fermi and Bose gas mixtures 
\cite{Pitaevskii_Stringari_book,Pethick_Smith_book}, as well as lattice models, such as Bose and Fermi Hubbard models \cite{Pitaevskii_Stringari_book,Pethick_Smith_book}, and Heisenberg-like models of interacting spins \cite{Sachdev_2011} 
where the role of the correlation function $\overline{G^{(2)}}$ is taken by the neighboring spin-spin correlation function \cite{Rist}. We also note that these thermodynamic relations are similar to those that have been derived in the context of Tan's contact parameter and the related Tan thermodynamic relations  \cite{Tan_contact_2008,Werner2009,Zwerger_Tan_2011,Braaten_Tan_review}; see, e.g., the treatments summarized in Chap. 18.3 of Ref. \cite{Pitaevskii_Stringari_book} and in Ref. \cite{Cherny2021}, which we closely followed here. In retrospect this is not surprising, because Tan's contact,  which characterizes the strength of the tails of the momentum distribution of an ultracold atomic gas, is known to be directly proportional to the (spatial) local atom-atom pair correlation function $g^{(2)}$ \cite{Zwerger_Tan_2011,Minguzzi_2013,De_Rosi_2023,kerr2024analytic}. We emphasize, however, that deriving the thermodynamic and Maxwell relations presented in this Letter does not rely on, and does not require the knowledge of, Tan's contact and Tan's thermodynamic relations.

~

\emph{\textbf{Appendix B: The stochastic projected Gross-Pitaevskii approach}}---In the SPGPE approach \cite{castin2000,blakie2008dynamics,spgpe_Rooney_2012}, the quantum field operator $\hat{\Psi}(x,t)$ is decomposed into two regions,
a $c$-field region and an incoherent thermal region. The $c$-field region
contains highly occupied low-energy modes and is described
by a single complex-valued classical field $\Psi_{\mathbf{C}}(x,t)$. The incoherent region, on the other hand, 
contains sparsely occupied high-energy modes that act as
an effective thermal bath, treated as static, with temperature
$T$ and chemical potential $\mu$ that governs the thermal average
number of particles in the system (in the $c$-field region). The boundary between these two regions is defined by an appropriately chosen energy cutoff $\epsilon_{\text{cut}}$ \cite{SPGPE_cutoff}.

In this approach, the thermal equilibrium state of the system is prepared by evolving the simple
growth SPGPE
 for the complex $c$-field $\Psi_{\mathbf{C}}(x,t)$ \cite{blakie2008dynamics,spgpe_Rooney_2012},
\begin{align}
	\label{eq:breathing_SPGPE}
	d\Psi_{\mathbf{C}}&(x,t)=\mathcal{P}^{(\mathbf{C})}\!\left\{ - \frac{i}{\hbar}\mathcal{L}_{0}^{(\mathbf{C})}\Psi_{\mathbf{C}}(x,t)dt 
	\right. \nonumber\\
	&\left.
	+ \frac{\Gamma}{k_BT}(\mu-\mathcal{L}_{0}^{(\mathbf{C})})\Psi_{\mathbf{C}}(x,t)\, dt + dW_{\Gamma}(x,t)\!\right\}.
\end{align}
Here, the projection operator $\mathcal{P}^{(\mathbf{C})}\{\cdot\}$ sets up the high-energy cutoff 
$\epsilon_{\mathrm{cut}}$, whereas $\Gamma$ is the so-called growth rate responsible for the coupling between the $c$-field and the effective reservoir (served by the incoherent region). In addition, $\mathcal{L}_{0}^{(\mathbf{C})}$ is the Gross-Pitaevskii operator defined by
\begin{equation}
	\mathcal{L}_{0}^{(\mathbf{C})} = -\frac{\hbar^{2}}{2m}\frac{\partial^{2}}{\partial x^{2}} + V(x,t) + \chi|\Psi_{\mathbf{C}}(x,t)|^{2}, 
\end{equation}
where $V(x,t)$ is the external trapping potential, if any. The last term, $dW_{\Gamma}(x,t)$, in Eq.~\eqref{eq:breathing_SPGPE} is a complex-valued stochastic white noise term with the following nonzero correlation:
\begin{equation}
	\langle dW_{\Gamma}^{*}(x,t)dW_{\Gamma}(x',t) \rangle = 2\Gamma\delta(x-x')dt.
\end{equation}

As we mentioned in the main text, the stochastic realizations of the $c$-field $\Psi_{\mathbf{C}}(x,t)$ prepared via the SPGPE after a sufficiently long evolution time sample the grand-canonical ensemble of thermal equilibrium states of the system. These stochastic realizations can then be evolved in real time according to the mean-field projected Gross-Pitaevskii equation \cite{blakie2008dynamics}, following a certain out-of-equilibrium protocol. This would then represent real-time dynamical evolution of the system starting from an initial thermal equilibrium state. In this Letter, however, we do not simulate any real-time dynamics; instead, we are interested in the pair correlation function $g^{(2)}$ of a 1D quasicondensate at thermal equilibrium. Accordingly, our simulations involve only the SPGPE stage.

Denoting the SPGPE realizations of the complex $c$-field $\Psi_{\mathbf{C}}(x,t)$ after a sufficiently long evolution time via $\Psi_{\mathbf{C}}(x)$, the thermal equilibrium values of physical observables are then calculated in terms expectation values of products of $\Psi_{\mathbf{C}}(x)$ and its complex conjugate $\Psi_{\mathbf{C}}^{*}(x)$ 
over a large number of stochastic realizations. This is much in the same way as calculating the same observables in terms of expectation values over normally ordered products of quantum field operators $\hat{\Psi}(x)$ and $\hat{\Psi}^{\dagger}(x)$, except that their noncommuting nature is ignored.
 As an example, the particle number density $n(x) = \langle \hat{\Psi}^{\dagger}(x) \hat{\Psi}(x) \rangle$ in the SPGPE approach is calculated as  $n(x)  = \langle \Psi_{\mathbf{C}}^{*}(x) \Psi_{\mathbf{C}}(x) \rangle$, where the brackets $\langle{...}\rangle$ refer to ensemble averaging over a large number of stochastic trajectories; similarly the pair correlation function $g^{(2)}$ can be computed via Eq.~\eqref{g2-SPGPE} of the main text.


\begin{thebibliography}{57}
\bibitem [{ext()}]{extensive}%
  \BibitemOpen
  \href@noop {} {}\bibinfo {howpublished} {The extensive nature of
  $\overline{G^{(2)}}$ follows immediately from the fact that $\chi
  \overline{G^{(2)}}/2$ (where $\chi$ is intensive) is simply the expectation
  value of the interaction part of the Hamiltonian, $\langle
  \hat{H}_{\mathrm{int}}\rangle=\chi \overline{G^{(2)}}/2$, which itself must
  be extensive because the overall Hamiltonian energy and the kinetic energy
  part alone are also extensive. For a uniform system of length $L$, this can
  also be seen explicitly through $\overline{G^{(2)}}=L\langle
  \hat{\Psi}^{\dagger}\hat{\Psi}^{\dagger}\hat{\Psi}\hat{\Psi}\rangle=Ln^2g^{(2)}$,
  where $L$ is extensive whereas $n$ and $g^{(2)}$ are intensive.}\BibitemShut
  {Stop}%
\bibitem [{alt()}]{alt_chi}%
  \BibitemOpen
  \href@noop {} {}\bibinfo {howpublished} {We note that one can also show
  within the standard microcanonical formalism of statistical mechanics that
  $\chi$ can be alternatively calculated from $S$ using $\chi
  =\!T\left(\partial S/\partial(\overline{G^{(2)}}\!/2)\right)_{U,L,N}$, which
  is analogous to showing that $P=-(\partial U/\partial L)_{S,N}=T(\partial S/
  \partial L)_{U,N}$.}\BibitemShut {Stop}%
\bibitem [{SPG()}]{SPGPE_cutoff}%
  \BibitemOpen
  \href@noop {} {}\bibinfo {howpublished} {For results presented in Figs.~1 and
  2 of the main text, the energy cutoff $\epsilon_{\text{cut}}$ was chosen such
  that the occupancy of the highest energy modes in the plane wave basis was
  approximately 1.}\BibitemShut {Stop}%
\end{thebibliography}

%

\end{document}